\documentclass[galaxies,article,accept,oneauthor,pdftex]{Definitions/mdpi} 

\firstpage{1} 
\pubvolume{xx}
\issuenum{1}
\articlenumber{5}
\pubyear{2021}
\copyrightyear{2021}
\history{Received: date; Accepted: date; Published: date}
\Title{Big and young supermassive black holes in the early Universe}

\Author{Tullia Sbarrato$^{1}$\orcidA{}}

\AuthorNames{Tullia Sbarrato}

\address[1]{%
$^{1}$ \quad INAF -- Osservatorio Astronomico di Brera, via E.\ Bianchi 46, I--23807 Merate (LC), Italy} 

\corres{Correspondence: tullia.sbarrato@inaf.it}

\abstract{
Blazars are Active Galactic Nuclei characterized by relativistic jets launched in the vicinity of the central engine (i.e.\ a supermassive black hole), that are oriented close to our line of sight.
Their peculiar orientation makes them very efficient tracers of the overall jetted population, 
and due  to their brightness they can be visible up to very high redshifts. 
A deep knowledge of these objects can provide fundamental clues to the models of formation and growth of the first supermassive black holes, 
but their search in the early Universe must be careful and follow a systematic approach. 
The discovery in the last $\sim15$ years of extremely massive blazars at very high redshifts ($M_{\rm BH}>10^9M_\odot, \, z>4$) revolutionized our perception of their earliest evolution: there seem to be different formation epochs for extremely massive black holes hosted in jetted ($z\sim4$) and non–jetted systems ($z\sim2.5$). 
This is not easy to explain, since one would expect that jetted sources accrete less efficiently. 
Small differences in the population are also derived from the search of such high--$z$ sources: we will go through the open questions in order to understand where the common knowledge stands and which steps must be undertaken to better understand the formation and common evolution of supermassive black holes and jets in the early Universe.
}


\begin{document}
%%%%%%%%%%%%%%%%%%%%%%%%%%%%%%%%%%%%%%%%%%

%%%%%%%%%%%%%%%%%%%%%%%%%%%%%%%%%%%%%%%%%%

\section{Introduction}

Active Galactic Nuclei (AGNs) are accreting supermassive black holes located in the centre of massive galaxies. 
They are characterized by a strong persistent emission covering the whole electromagnetic spectrum. 
In particular, optical--UV radiation is produced by the release of gravitational energy from the matter accreting towards the central black hole. 

About 10\% of the overall AGN population show collimated relativistic jets launched from the vicinity of the central engine  \cite{urry95}. 
When reaching the intra-cluster medium, these structures are slowed down and their release of energy create extended lobes, filled with strong magnetic fields, that allow them to emit isotropically in the radio band through syncrotron cooling. 
For this reason, an AGN showing a strong radio emission is generally thought to be hosting a jet. 
A quantitative indication of the jet presence was introduced by \citet{fanaroffriley74} in their effort to classify the position of bright radio regions of extragalactic sources: the radio--loudness $R=F_{\rm r}/F_{\rm opt}$ measures the dominance of the radio over optical monochromatic emissions (i.e.\ the presence of a jet brighter than the accretion emission). 
Being emitted from extended lobes, radio emission is visible independently of the jet orientation with respect to an observer line of sight. 
Collimated jets instead emit at all wavelengths, and are strongly boosted due to relativistic beaming along their propagation direction. 
Their emission can thus dominate the whole AGN spectrum when aligned along an observer line of sight, in which case the sources are classified as blazars. 
The collimated radio emission from the jet is easily much brighter than the isotropic lobe radio emission. 
The radio-loudness will thus reach extreme values of few thousands in the case of a well aligned jet and will be a first indicator of alignment. 

Blazars broad band emission is characterized by two humps: syncrotron emission is responsible for the one peaking around sub-millimetre to soft X-rays \cite{padovani17}, while inverse Compton on electrons or hadrons produce the high-energy peak, that falls between X-rays and TeV $\gamma$-rays. 
This peculiar emission profile makes blazars easily detectable at very high--frequencies.
In fact, quasars detected in the $\gamma$--rays are generally blazars: AGN all-sky catalogs such as the one compiled by {\it Fermi}/LAT are mainly composed by such sources (see Section \ref{subsec:surv}).

%%%%%%%%%%%%%%%%%%%%%%%%%%%%%%%%%%%%%%%%%%
\section{Finding high--redshift blazars}

The peculiar orientation of blazars make them the perfect tracers for the whole family of jetted AGNs. 
For each blazar observed and studied, one can assume the presence of many analogous jetted AGNs 
with their jets oriented in other directions and same features (black hole mass, accretion rate, jet power, ... ). 
Knowing  the viewing angle ($\theta_{\rm v}$) and the jet beaming angle ($\theta_{\rm b}$) 
of a blazar allows us to estimate the number of jetted quasars composing its parent population ($N_{\rm parent}$). 
If an object is observed under a viewing angle smaller than the beaming angle, in fact, 
one can estimate the number of objects that have the same properties but jets pointing elsewhere. 
%This estimate is set by the ratio between the all sky surface and the solid angle subtended by the jet beaming angle. 
%This approach is qualitatively schematized in 2D in Figure , but a quantitative definition is needed.

We can assume that the emitting region is relativistically moving with an overall Lorentz factor 
$\Gamma = 1/\sqrt{1-\beta^2}$, and emitting isotropically in its own rest frame. 
According to an outside observer 
the region appears to emit half of its photons into a cone with semi–aperture $\theta_{\rm b}$, 
where $\sin\theta_{\rm b}=1/\Gamma$ \cite{rybicki85}. 
The emitting region is highly relativistic, and therefore moves with $\beta=v/c\sim1$.
This implies a very small value for $\theta_{\rm b}$, and therefore: 
\begin{equation}
    \sin\theta_{\rm b} \sim \theta_{\rm b} \sim \frac{1}{\Gamma}
\end{equation}
We therefore define the beaming angle of a relativistic jet as $\theta_{\rm b}=1/\Gamma$.

We can then derive the ratio between the number of objects observed with $\theta_{\rm v}\leq1/\Gamma$ 
and the total number of analogous jetted AGNs, but randomly oriented. 
This estimate corresponds to the ratio $f_{\Omega}$ between the solid angle subtended 
by the jet beaming angle $\Omega_{\rm b}$ and the all sky surface $\Omega_{4\pi}$. 
Note that the jetted emission from a single AGN is formed by a jet (that we actually see in the case of a blazar) and a counterjet. 
Therefore, we must take into account both jet and counterjet in the calculation, and we will be able to 
express $f_\Omega$ only as a function of the Lorentz factor $\Gamma$:
\begin{equation}
    f_\Omega = \frac{2\Omega_{\rm b}}{\Omega_{4\pi}} = 1-\beta \simeq \frac{1-\beta^2}{2} = \frac{1}{2\Gamma^2}
\end{equation}
That corresponds to the ratio between the number of objects seen at $\theta_{\rm v} \leq 1/\Gamma$ 
and the total number of jetted objects with jets oriented randomly, i.e. $N_{\rm parent} = 1/f_\Omega$. 
Therefore, the number of analogous jetted AGNs oriented in the others directions is:
\begin{equation}
    N_{\rm parent} \simeq 2\Gamma^2 = 338 \left( \frac{\Gamma}{13} \right)^2
\end{equation}
where $\Gamma=13$ is a typical value for a blazar \cite{ghisellini09}. 
At low redshift this is not particularly more efficient than compiling a complete catalog 
of misaligned AGNs to explore the parent population, since extensive studies and complete 
quasar catalogs are actually available. 
At high redshift, instead, a complete collection of jetted AGNs is not yet available, 
and therefore \textbf{blazars can be main actors in the systematic study of the jetted AGN population}, 
thanks to their statistical relevancy.

In the 2000s, a few blazars where discovered serendipitously at $z>4$ by observing their high--energy components. 
The most notable example is Q0906+6930 \cite{romani04,romani06}, the only high--redshift blazar ever detected by EGRET 
and for 16 years the most distant blazar known, being located at $z=5.47$.
Its blazar nature was further confirmed with VLBI, VLA and {\it Chandra} observations. 
The rest of the $z>4$ blazar serendipitously discovered between 2000 and 2010 are consistently very radio--loud, 
and their blazar nature could be confirmed through their X--ray emission \cite{bassett04,shemmer05,young09}.
No $z>4$ source other than Q0906+6930 has been serendipitously discovered in the GeV frequecy range. 

%%%%%%%%%%%%%%%%%%%%%%%%%%%%%%%%%%%%%%%%%%

\subsection{All--sky surveys}
\unskip
\label{subsec:surv}

%The peculiar orientation of blazars, combined with the strong boosting of their jets, makes blazars 
%extremely luminous, and therefore they should be in principle observable even at very high redshift. 
%In principle, then, they should be easy to observe and extremely useful for high–redshift statistics. 
%Nevertheless, only few of them are known at $z > 4$, and all of them were discovered serendipitously.

The systematic discovery of blazars is guaranteed at low--redshift by high--energy instruments like {\it Fermi}/LAT. 
This instrument provides the largest $\gamma$–ray catalog of AGNs (the Third Catalog of AGNs detected by Fermi, 3LAC \cite{ackermann15}), 
that mostly include blazars, along with very few radio--galaxies and unclassified AGNs. 
3LAC includes 414 FSRQs and 604 BL Lacs, but they all have redshift $z\leq3.1$. 

This cut is not due only to the sensitivity limit, though, but the high redshift and the blazar sequence are mostly responsible for it.
At high redshift, in fact, we expect to see only the most luminous objects, and the most 
luminous blazars have  SED peaks at smaller frequencies \cite{fossati98,ghisellini17}. 
This effect, along with the high redshift, shifts the high--energy hump of the SED below 100 MeV. 
The detection of this component by the {\it Fermi}/LAT becomes therefore unlikely. 
Instead, a powerful blazar can be more easily detected at frequencies just before the inverse Compton peak 
by hard X--ray instruments, like the Burst Alert Telescope (BAT) onboard the Swift satellite \cite{gehrels04}. 
This is true already at $z\sim3$. 
Unfortunately the {\it Swift}/BAT has a very high sensitivity limit, and therefore is extremely limited to objects up to $z\sim 4$. 
At redshift higher than 4, a systematic search at high energies is therefore impossible to perform. A different approach is needed.

There are only 5 blazars known at $z>3.1$ detected by {\it Fermi}/LAT, that have been actively searched in the all--sky data of the instrument \cite{ackermann17}.
The discovery of their $\gamma$--ray emission follows a systematic and specific approach, starting from known high--redshift radio--loud quasars and digging into $\sim92$ months of {\it Fermi} Pass 8 source class photons. 
A single source from this sample is located at $z>4$: NVSS J151002+570243 ($z=4.3$) is currently the farthest known $\gamma$--ray emitting blazar, but it could not be discovered simply by the all--sky {\it Fermi}/LAT survey, because of the limited sensitivity.

\subsection{Candidate selection}

Serendipitous findings of high--redshift blazars is clearly not efficient to have a complete census of these sources.
For this reason, a systematic approach has been followed in the last years. 
As already pointed out, very bright blazars have been observed to have their signature humps shifted towards lower frequencies \cite{fossati98,ghisellini17}. 
This effect, described by the so--called blazar sequence, only affects the jet emission. 
All the thermal radiation originated in the accretion system is thus left visible in the rest--frame UV--optical frequency range. 
This feature allowed different research groups to select reliable blazar candidates. 
A systematic approach is possible if a complete quasar catalog is used as a starting point for searching for blazar candidates. 
This is feasible since only FSRQs can be observed at such high redshift: BL Lacs (if present) would not have reliable redshift estimates. 
Specifically in the optical range, therefore, we can be sure that the thermal emission from the accretion disc dominate the continuum emission, 
along with the emission lines from broad and narrow line regions that are visible in that wavelength range.

In order to select good high--redshift blazar candidates, one should limit the search to  quasars with $z > 4$ if 
starting from an optical spectroscopic quasar catalog. 
It is possible though to select from scratch new high--$z$ quasars, as was done in order to discover the farthest blazar currently known, PSO J0309+27 \cite{belladitta20}. 
These authors selected bright and compact $i$--dropout sources from Pan-STARSS survey, with a relatively strong radio--counterpart. 
Their selection from the near-IR--optical survey is the standard procedure to find new $z>5.7$ quasars, followed by various groups focusing on the matter. 
Their selection in radio, instead, mirrors the search for extreme radio--loudness: at $z>4$ the radio flux is bright enough to allow cuts towards extreme radio--loudness values (namely $>100$), while at higher redshift a strong radio flux is already a reliable indication of the presence of a jet. 

This more systematic approach has currently lead to the discovery and classification of 15 blazars at $z>4$, from well defined, flux--limited samples, that allow for the derivation of strong contraints on the overall jetted population (see the following Sections).

%%%%%%%%%%%%%%%%%%%%%%%%%%%%%%%%%%%%%%%%%%
\section{Discussion}

The discovery of the high--redshift blazar population is still an open process, but the sample currently available can already tell us something about the massive black holes accreting in the early Universe. 
All results need to be considered as a lower limit on the population, since the complete census is still in the process of being completed. 

\subsection{How many jetted AGNs at high--$z$?}
\label{sec:com_num_den}

We already demonstrated in detail how blazars are a statistically significant representation of the jetted population, because of their orientation. 
This peculiarity can be used to compare the jetted population with their non--jetted counterparts. 
We focus on the most massive sources, i.e.\ $M_{\rm BH}>10^9M_\odot$, to explore the evolution of the most extreme supermassive black holes across the cosmic time. Moreover, most quasars observed at very high--redshift with a measured mass host such a massive black hole. Henceforth, we can draw interesting conclusions on the AGN population that we are able to observe in the early Universe. 

Blazars selected from complete optical quasar surveys are generally limited to a specific area of the sky. 
Those systematically selected at $z>4$ from the SDSS and FIRST survey, in particular, are limited to the solid angle $\Omega_{\rm SDSS+FIRST}\simeq8770{\rm deg}^2$. 
It is therefore reasonable to assume that the amount of blazars we find in that specific sky region is a fraction $\Omega_{\rm SDSS+FIRST}/4\pi$ of the total number of blazars with those features distributed in the whole sky. 
Knowing the size of the selection area of the blazar candidates allows their discovery to be more statistically relevant than serendipitously discovered sources. 
The observation of one of them traces the presence of $N_{\rm parent}$ jetted AGNs, where 
\begin{equation}
    N_{\rm parent} \simeq 2\Gamma^2 \frac{4\pi}{\Omega_{\rm SDSS+FIRST}} = 2\Gamma^2 \frac{41253 {\rm deg}^2}{8770 {\rm deg}^2}.
\end{equation}
It is therefore possible to calculate how many jetted AGNs with same black hole mass and accretion rate can be inferred.
This can thus be used efficiently to study the distribution of early extremely massive black holes, providing constraints for models of supermassive black hole formation and evolution.

\begin{figure}[!h]
\centering
\includegraphics[width=\textwidth]{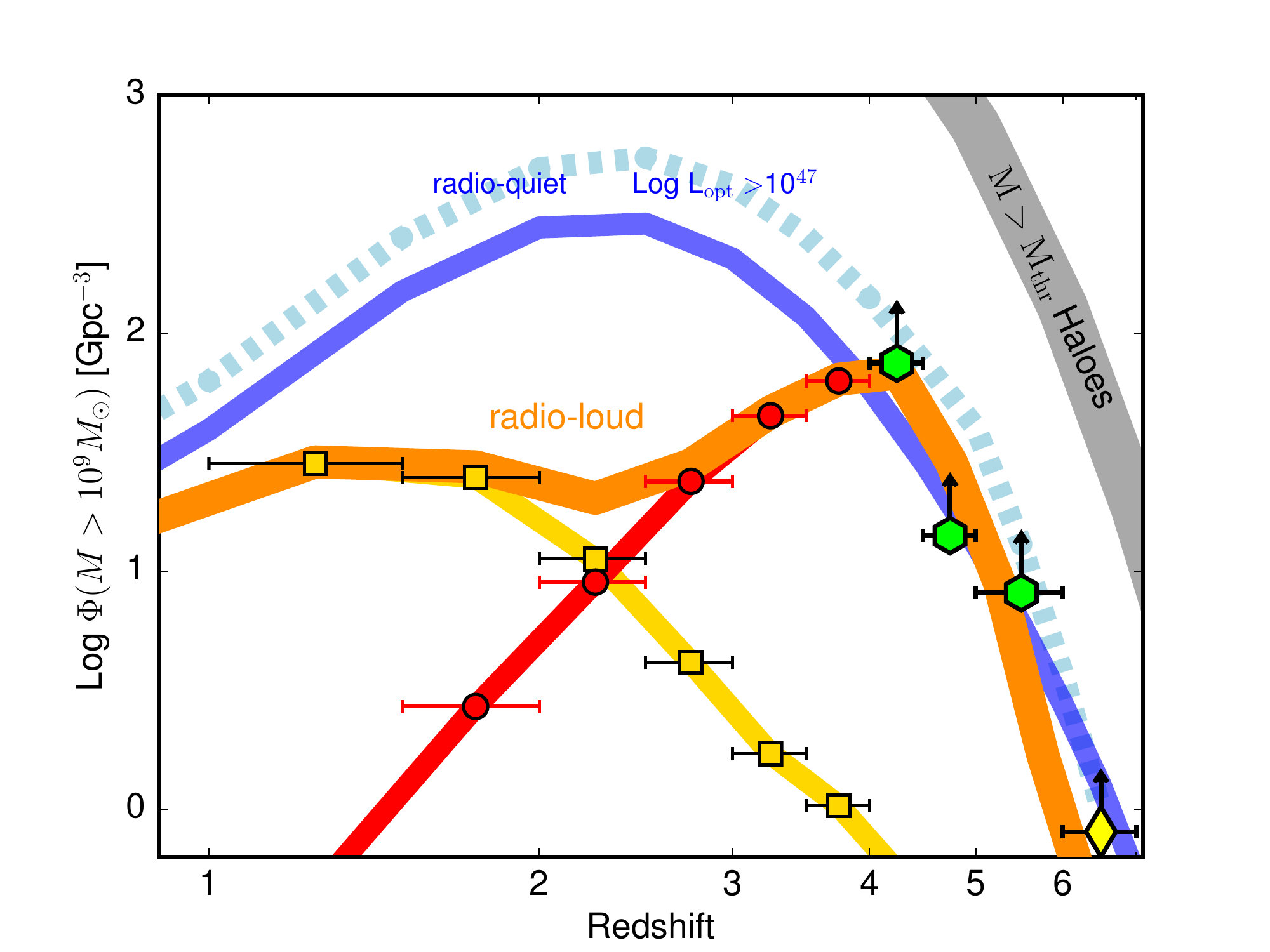}
\caption{Comoving number density of active  supermassive black holes with $M_{\rm BH}\geq10^9M_\odot$ hosted in jetted (orange line) and non--jetted AGNs (blue line). 
         The non--jetted density is derived from the luminosity function of radio--quiet quasars in SDSS \cite{shen20}, by assuming that all quasars with $L_{\rm opt}>10^{47}{\rm erg/cm^2/s}$ host black holes with masses we are interested in. 
         As a comparison, the light blue dashed line is the comoving number density derived from \citet{hopkins07}, used in previous versions of this figure. 
         The jetted--related curve is instead formed by the sum of luminosity functions from {\it Fermi}/LAT (yellow line and points) and {\it Swift}/BAT (red line and points), up to $z\simeq4$. 
         Green hexagonal lower limits are derived from the systematic classification of blazar candidates from SDSS+FIRST, based on X--ray observations and corrected by the sky area covered by the survey. 
         The yellow diamond lower limit refers to PSO J0309+27, the farthest blazar currently known.
        }
\label{fig:phi_z}
\end{figure}   

Figure \ref{fig:phi_z} shows the comoving number density of extremely massive black holes (i.e. with $M_{\rm BH} > 10^9M_\odot$) hosted by radio--quiet (blue line, derived from \citep{shen20}\footnote{
The comoving number density shown here was derived following the assumptions considered by \citet{ghisellini10} while extracting the same quantities from \citet{hopkins07}.
}) and radio--loud AGNs, i.e. non--jetted and jetted sources. 
On the other hand, the comoving number density of radio--loud quasars is derived from {\it Fermi}/LAT \cite{ajello12} and {\it Swift}/BAT \cite{ajello09} blazar luminosity functions, assuming that every blazar traces the existence of $2\Gamma^2$ jetted quasars \cite{ghisellini10}. The blazar surveys of {\it Swift}/BAT and {\it Fermi}/LAT are limited at $z<4$,  thus at $z>4$ there is no complete survey of blazars that provides a reliable number density. For $z>4$ the density was previously assumed to decrease exponentially, as the corresponding one for radio–quiet objects.

Nevertheless, a hint of different density distributions between jetted and non–jetted objects was already visible in \cite{ghisellini10} and \cite{ghisellini13}. 
While radio--quiet objects seem to be more numerous at redshift $z \sim 2 - 2.5$, the distribution of jetted quasars suggests that they do not show such a peak, but more likely a plateau that extends at least up to $z \sim 4$, or maybe a second peak at $z \sim 4$. 
Those papers did not reach a firm conclusion about the real distribution of jetted AGNs at $z > 4$, lacking a significant number of objects in that redshift frame. In other words, they could not disentangle between a plateau and a high--redshift peak.

The number densities derived by classifying systematically selected blazar candidates are instead very significant. Current observations in fact clearly push towards an interesting conclusion: the density of extremely massive black holes hosted in jetted systems peaks at least around $z \sim 4$, while the non--jetted systems peak at $z \sim 2-2.5$. In other words, {\bf the number of heavy black holes hosted in jetted systems increases with cosmic time}, and peaks at $z > 4$. This suggests two different epochs of SMBH formation. The most striking feature of this result is that black holes that grow while developing a jet seem to be born earlier, or to grow faster, than black holes with same masses that do not show jets.

\subsection{Early quasar evolution}
\label{sec:quas_evol}

The presence of a jet in AGNs is commonly linked to high values of black hole spin. The black hole rotational energy, in fact, is a possible source of power to launch the jet \cite{blandford77,tchekhovskoy11}, while the angular momentum of the black hole itself easily identifies its launching direction. 
This does not facilitate a fast accretion, according to common knowledge. Rapidly spinning black holes, in fact, are generally associated to high accretion--based radiative efficiencies, since the innermost stable orbit of the accretion disc approaches the gravitational radius. \citet{thorne74} showed that the maximum spin that a black hole can reach through accretion is $a = 0.998$ (instead of $a = 1$), since the hole slows down by capturing accretion disc radiation, that exerts a counteracting torque. The accretion efficiency that corresponds to this spin value
is $\eta = 0.3$, larger than the efficiency in the case of a non--rotating black hole ($\eta = 0.08$). 
A larger efficiency corresponds to a lower overall accretion rate (since $\dot M = L_{\rm d}/\eta c^2$). 
For a fixed luminosity, then, a black hole accretes less matter and grows slower. 
A spinning black hole accreting at Eddington rate (i.e.\ at the highest rate allowed for a standard
disc), if we assume that it doubles its mass on a \citet{salpeter64} time, would need 3.1 Gyr to grow from a seed of $100 M_\odot$ to $10^9M_\odot$ \cite{ghisellini13}. 
Hence, such massive black holes would not be visible at $z > 2.1$, while we see from Figure \ref{fig:phi_z} that their preferential formation epoch seems to be around $z \sim 4$.

\begin{figure}[!h]
\centering
\includegraphics[width=\textwidth]{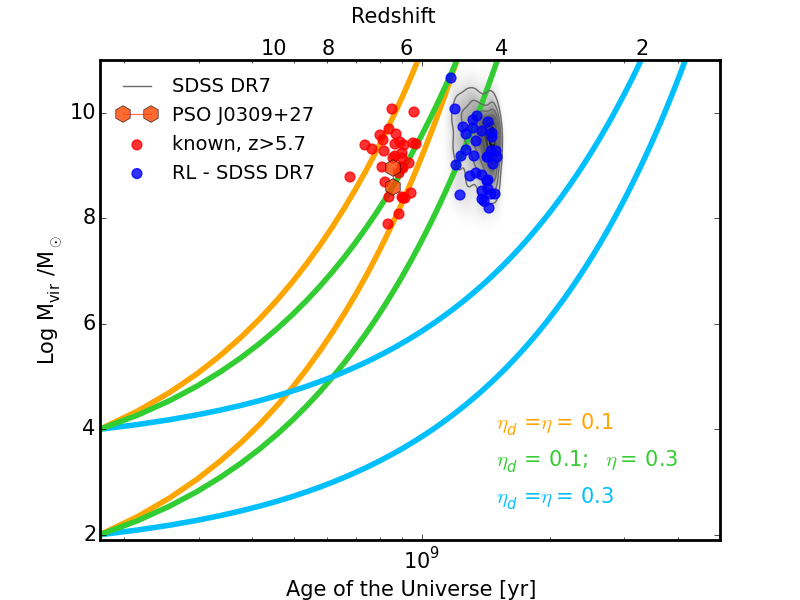}
\caption{Black hole mass evolution as a function of the age of the Universe. Red points are the known quasars with virial--based mass measures at $z>5.7$. 
         The hexagons show the mass range of the highest--$z$ blazar known \cite{belladitta20}. 
         The grey contours show the whole SDSS DR7 quasar sample, including the few tens of radio--loud sources (blue points). 
         The solid lines show the Eddington--limited evolution based on the Salpeter timescale, starting from two different black hole seeds, and assuming that (i) the energy released during the accretion $\eta$ coincides with the accretion disc radiative efficiency $\eta_{\rm d}$ in the case of a Shakura \& Sunyaev disk around a non--spinning black hole (yellow line), (ii) same but in the case of a maximally spinning black hole (green line), (iii) the black hole is maximally spinning and thus the energy release is $\eta=0.3$, but only a fraction of it is radiated through the accretion disc (light blue line).}
\label{fig:salpeter}
\end{figure}   

Some options for a faster accretion in presence of a jet are explored in literature \cite{ghisellini13,mazzucchelli17}. 
A fascinating toy model suggests how the jet could facilitate a fast accretion, contrary to what is generally thought. 
The available energy, in fact, is not all radiated away, but it may contribute in amplifying the magnetic field and thus launching the jet. 
We can define two different contributions to the global efficiency $\eta$: one related to the disc luminosity (i.e.\ the radiative efficiency $\eta_{\rm d}$), and one implied in generating the magnetic field that helps in launching the jet $\eta_{\rm j}$:
\begin{equation}
    \eta = \eta_{\rm d} + \eta_{\rm j}.
\end{equation}
In this case, to produce a fixed luminosity, we need a larger accretion rate, since part of the gravitational energy is spent for some non radiative processes necessary to form the jet. 
In other words, a spinning black hole can accrete faster if some of its accretion efficiency is spent for the jet ($\eta_{\rm j} > 0$ and $\eta = 0.3$).
Considering this, the accretion is faster, but black holes with $M_{\rm BH} > 10^9 M_\odot$ are still hard to form before $z \sim 4 - 5$.

The blazars found at $z \geq 5$ ($\sim$1Gyr) all have $M_{\rm BH} \geq 10^9 M_\odot$. 
If this black hole is maximally spinning ($\eta = 0.3$), its mere presence requires $\eta_{\rm j} \geq 0.2$ and $\eta_{\rm d} \leq 0.1$, as can be seen in Figure \ref{fig:salpeter}, if we assume a black hole seed of $100M_\odot$. 
Assuming a more massive seed helps, but we are still forced to consider a maximally spinning black hole that does not convert its entire gravitational energy into radiation, but also in generating a magnetic field to launch the jet. 
This is reasonable and (probably) necessary for high--redshift objects, but is not required to justify the presence of heavy black holes in jetted systems at low redshift. 
At $z \sim 2$, in fact, more time is available to form such massive black holes. 
Note also that a fast accretion requires a large amount of cold gas in the surroundings of the black hole, to continuously fuel the accretion. 
This is a reasonable picture at high redshift, but the black hole environment at lower redshift is generally poorer. 
This means that other, less extreme and more plausible processes can drive supermassive black hole formation and growth. 
This is the reason why the presence of such massive objects is not a major problem, while their presence at z$ > 4$ is harder to be justified.

At this stage, the toy model here described is only a plausible idea to justify such a relevant population of heavy black holes in jetted systems, and therefore the presence of these sources at high redshift still remains an interesting open issue.
Another straightforward option is that these supermassive black holes are accreting in a super--Eddington regime, but there is no observation--based indication that this could be the case for the currently known sources.

\section{Open issues}

\subsection{Missing jetted sources}
\label{sec:missing}

Figure \ref{fig:phi_z} clearly shows another peculiar behaviour of high--redshift jetted sources with respect to non--jetted ones and their lower redshift counterparts.
Not only radio--loud sources seem to be differently distributed, but at $z\geq4.5$ they seem almost to outnumber the radio--quiet ones, or at least to have comparable number densities. This is strikingly different from the ratio of the two populations at low redshift: it is calculated that jetted AGNs are $\sim10\%$ of the overall AGNs population, certainly not half, nor its entirety. 

A discrepancy in the populations ratio is already evident when considering only the jetted population. 
The systematic search of blazar candidates, as in \cite{sbarrato13}, leads to a blazar--to--radio--loud ratio very different from the lower redshift one: at least 8 blazars\footnote{
Other 3 sources will be classified as blazars in an upcoming paper, making the blazar--to--radio ratio even more extreme.
} 
have already been classified out of the sample of SDSS+FIRST spectroscopic quasar catalog, where radio--loud sources were only 53. 
This translates in a ratio of $\sim15\%$ blazars in a sample of radio--loud quasars at $z\geq4$, clearly much higher than what expected from the alignment--based calculations, i.e.\ less than $1\%$ \cite{ghisellini16}.
Might misaligned jetted sources be missing? Or not visible?

Clearly, the discrepancy could be due to different jet features at very high redshift, that would twist the ratio between aligned and misaligned jetted sources from the standard $2\Gamma^2$. 
This is rather hard to justify, though, since the broad--band SED profile of $z>4$ blazars does not show major differences when compared to lower redshift ones. 
A known effect is the quenching of extended radio emission in favor of up--scattering of CMB photons by the lobe electron population happening at $z>3$ \cite{ghisellini14}. 
The CMB energy density at such high redshifts becomes comparable with the magnetic energy density in lobe regions of jetted AGNs, offering an alternative and preferential interaction to electrons that normally would emit by synchrotron cooling. 
This effect is widely confirmed by the observation of extended X--ray emissions visible along jet directions in high--$z$ quasars \cite{yuan03} and easily justifies the loss of widely misaligned blazar counterparts at $z>4$. 

Numbers suggest instead that even the mildly misaligned sources are missing.
This discrepancy can be fixed by taking into consideration the larger IR luminosity observed in high--$z$ quasars, that has been observed in $z>7$ quasars and partially obscured $z>4$ ones \cite{mortlock11,fan20}: high--redshift AGNs might be over--obscured by ``bubbles'' of dust, possibly typical in the early Universe \cite{ghisellini16}.
Such bubbles would be able to completely obscure the nuclear emission, up to a threshold luminosity that would eventually eject the dust because of its strong radiation pressure. According to this scenario, only the jet would be able to pierce through the bubble, opening a small aperture that would allow to directly observe the thermal emission only when the jet is closely aligned to the line--of--sight.
This would prevent slightly misaligned jetted AGNs from being visible as standard quasars in optical--UV photometric and spectroscopic catalogs during the over--obscured phase. 

These features reconciles the results in the frame of jetted sources, but do not help in understanding why radio--loud (jetted) quasars are as or more numerous than radio--quiet (non--jetted) quasars. \emph{Might the presence of a jet actually help in fast accretion?}

\subsection{Where does the jet point?}
\label{sec:differences}

Many sources classified as blazars on the basis of X--ray observations from the $z\geq4$ SDSS+FIRST sample were also observed with VLA or VLBI high--resolution configurations. 
while in most cases the two approaches yield to consistent results \cite{frey15}, some of these sources do not show the radio features expected from blazar–like AGNs \cite{cao17}.
As an example, some show extended radio--emitting regions that suggest larger viewing angles, or a strong core--dominance is lacking as in the case of PSO J0309+27 \cite{spingola20}.

Different possibilities could be invoked to handle this interesting inconsistency. 
It is possible that low-- and high--redshift jets in AGNs emit with different processes, that still have to be completely explored. 
A more simple explanation relies on a different geometry of the jet: X--ray and radio emissions are emitted at different distances from the central engine, and while at low--$z$ the relativistic jet is clearly able to propagate along the same direction in the whole time, at very high--$z$ this might change. 
Different densities of the medium where the jet propagate might make its path difficult, and ultimately bend the jet. 
This would explain the different viewing angles derived from X--ray and radio emissions. 

This option clearly needs to be further explored, and it will constitute an interesting field to be investigated in the upcoming years, with more sensitive facilities and more detailed modeling of the early formation of supermassive black holes and relativistic jets.  
A tight collaboration between X--ray, radio and theoretical astronomers will be crucial, being clearly necessary a completely holistic approach to solve this puzzle.

%%%%%%%%%%%%%%%%%%%%%%%%%%%%%%%%%%%%%%%%%%
%\section{Conclusions}

%This section is not mandatory, but can be added to the manuscript if the discussion is unusually long or complex.

%%%%%%%%%%%%%%%%%%%%%%%%%%%%%%%%%%%%%%%%%%
\vspace{6pt} 

%%%%%%%%%%%%%%%%%%%%%%%%%%%%%%%%%%%%%%%%%%
%% optional
%\supplementary{The following are available online at \linksupplementary{s1}, Figure S1: title, Table S1: title, Video S1: title.}


%%%%%%%%%%%%%%%%%%%%%%%%%%%%%%%%%%%%%%%%%%
%\authorcontributions{For research articles with several authors, a short paragraph specifying their individual contributions must be provided. The following statements should be used ``conceptualization, X.X. and Y.Y.; methodology, X.X.; software, X.X.; validation, X.X., Y.Y. and Z.Z.; formal analysis, X.X.; investigation, X.X.; resources, X.X.; data curation, X.X.; writing--original draft preparation, X.X.; writing--review and editing, X.X.; visualization, X.X.; supervision, X.X.; project administration, X.X.; funding acquisition, Y.Y.'', please turn to the  \href{http://img.mdpi.org/data/contributor-role-instruction.pdf}{CRediT taxonomy} for the term explanation. Authorship must be limited to those who have contributed substantially to the work reported.}

%%%%%%%%%%%%%%%%%%%%%%%%%%%%%%%%%%%%%%%%%%
\funding{This research received no external funding.}
%Please add: ``This research received no external funding'' or ``This research was funded by NAME OF FUNDER grant number XXX.'' and  and ``The APC was funded by XXX''. Check carefully that the details given are accurate and use the standard spelling of funding agency names at \url{https://search.crossref.org/funding}, any errors may affect your future funding.}

%%%%%%%%%%%%%%%%%%%%%%%%%%%%%%%%%%%%%%%%%%
%\acknowledgments{In this section you can acknowledge any support given which is not covered by the author contribution or funding sections. This may include administrative and technical support, or donations in kind (e.g., materials used for experiments).}

%%%%%%%%%%%%%%%%%%%%%%%%%%%%%%%%%%%%%%%%%%
%% optional
\abbreviations{The following abbreviations are used in this manuscript:\\

\noindent 
\begin{tabular}{@{}ll}
AGN & Active Galactic Nucleus \\
BL Lac & BL Lacertae object \\
{\it Fermi}/LAT & The Large Area Telescope onboard the {\it Fermi Gamma-Ray Space Telescope} \\
FIRST & Faint Images of the Radio Sky at Twenty-Centimeters \\
FRSQ & Flat Spectrum Radio Quasars \\
SDSS & Sloan Digital Sky Surve \\
{\it Swift}/BAT & Burst Alert Telescope onboard the {\it Niel Gehrels Swift Observatory} \\
VLA & Very Large Array\\
VLBI & Very Long Baseline Interferometry\\
\end{tabular}}

%%%%%%%%%%%%%%%%%%%%%%%%%%%%%%%%%%%%%%%%%%
%=====================================
% References, variant B: external bibliography
%=====================================
\externalbibliography{yes}
\bibliography{sbarrato}

%%%%%%%%%%%%%%%%%%%%%%%%%%%%%%%%%%%%%%%%%%
\end{document}